\documentclass[aps,showpacs,prl,twocolumn,amsmath,amsfonts,amssymb,superscriptaddress]{revtex4}

\usepackage{graphicx}

\let\a=\alpha \let\b=\beta

\let\s=\sigma   
   \let\G=\Gamma
   
\let\Si=\Sigma   
   \let\io=\infty
 \def\ie{{i.e. }}

 \def\HH{{\cal H}}
\def\NN{{\cal N}} 
  
 \def\SS{{\cal S}}

\def\to{\rightarrow}  
 
\def\us{\underline\sigma}

\newcommand{\beq}{\begin{equation}} \newcommand{\eeq}{\end{equation}}
 
\newcommand{\Tr}{\text{Tr}}

\begin{document}

\title{
A solvable model of quantum random optimization problems
}

\author{Laura Foini}\affiliation{LPTENS, CNRS UMR 8549,
  associ\'ee \`a l'UPMC Paris 06, 24 Rue Lhomond, 75005 Paris,
  France.}
\affiliation{
SISSA and INFN, Sezione di Trieste, via Bonomea 265, I-34136 Trieste, Italy
}

\author{Guilhem Semerjian} \affiliation{LPTENS, CNRS UMR 8549,
  associ\'ee \`a l'UPMC Paris 06, 24 Rue Lhomond, 75005 Paris,
  France.}

\author{Francesco Zamponi}
\affiliation{LPTENS, CNRS UMR 8549, associ\'ee \`a l'UPMC Paris 06, 24
  Rue Lhomond, 75005 Paris, France.}


\begin{abstract}
We study the quantum version of a simplified model of optimization problems,
where quantum fluctuations are introduced by a transverse field acting on the 
qubits.
We find a complex low-energy spectrum of the quantum Hamiltonian, characterized
by an abrupt condensation transition and a continuum of level crossings
as a function of the transverse field.
We expect this complex structure to have deep consequences on the behavior
of quantum algorithms attempting to find solutions to these problems.
\end{abstract}

\maketitle

A large part of theoretical research in quantum computing has been
devoted to the development of algorithms that could use quantum
properties to perform computational
tasks faster than classical devices.  A typical problem that is
encountered in almost all branches of science is that of optimizing
irregularly shaped cost functions $H_P$: two standard examples are
$k$-SAT, $H_P$ counting the number of unsatisfied constraints on
$k$ boolean variables, and the coloring of a graph with $q$ colors
($q$-COL), $H_P$ then being the number of monochromatic edges.
The decision version (whether a solution, i.e. a configuration with $H_P=0$, 
exists) of both problems belongs to the class of NP-complete problems.
One is mostly interested in the scaling of the difficulty of these 
problems when the number $N$ of variables involved becomes large.
Besides the formal computational complexity theory which classifies 
the difficulty of problems according to a worst-case criterion,
their typical case complexity is often studied through random ensemble of 
instances, for instance assuming a flat distribution over
the choice of $M=\alpha N$ constraints on $N$ variables.

\begin{figure}
\includegraphics[width=.5\textwidth]{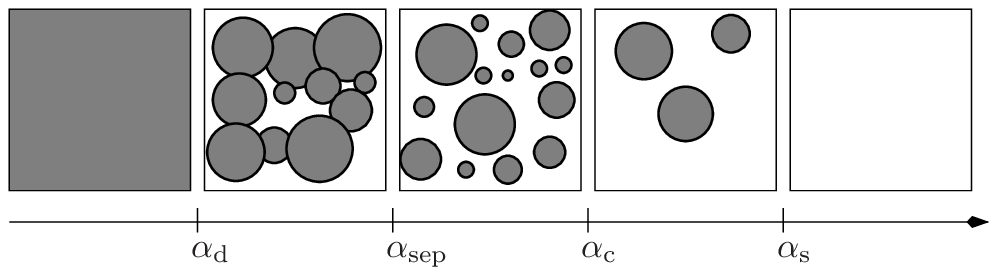}
\vskip-13pt
\caption{Pictorial representation of the different phase transitions
in the set of solutions of the Random Subcubes model~\cite{MZ08}.}
\vskip-18pt
\label{fig1}
\end{figure}

Statistical mechanics tools have provided a very
detailed and intricate picture of the configuration
space of such typical problem Hamiltonians $H_P$~\cite{CSP_classical}.
A key concept that emerged in this context is that of {\it clustering
of solutions}. The topology of the space of solutions changes abruptly 
upon increasing the density of constraints $\alpha$ in the following way:
{\it i)} at a first threshold $\a_{\rm d}$ it goes from a single connected 
cluster to a set of essentially disjoint clusters; 
{\it ii)} the number of clusters itself undergoes a transition at 
$\a_{\rm c} > \a_{\rm d}$ from a phase where it is exponential in $N$, thus 
defining an entropy of clusters (the {\it complexity}),
to a phase where the vast majority of solutions are contained in a finite
number of clusters;
{\it iii)} finally, the total entropy of solutions vanishes at 
$\a_{\rm s} > \a_{\rm c}$ where
the problem does not admit a solution anymore (the ground state energy becomes
positive).
This sequence of transitions (sketched in Fig.~\ref{fig1}) 
has a deep impact on the performances of most classical optimization algorithms,
that slow down dramatically deep in the clustered phase. 

To solve the optimization problem $H_P$ with a quantum computer,
one can choose its Hamiltonian to be 
$H = H_P + \G H_Q$, where $H_Q$
does not commute with $H_P$ and induces quantum fluctuations. One can switch
off $H_Q$ adiabatically in order to find the ground state of $H_P$~\cite{annealing}. 
The question that arises 
naturally is: what are the consequences
of the complicated structure of $H_P$ for the spectrum of $H$?
Indeed, we expect the spectrum to be strongly influenced by
the clustering phenomenon, and that this should have consequences on 
the performances of 
quantum algorithms as it does in the classical case. The influence of $H_P$ 
should be the strongest, and the easiest to analyze, for small $\G$.

Some initial steps in this analysis have been recently performed 
in~\cite{First}.
It was shown that increasing $\G$ can induce level crossings between classical 
ground states and classical low-energy excited states: in the thermodynamic 
limit these level crossings can be accompanied by exponentially small (in system size) gaps in 
the spectrum of $H$ that might be particularly dangerous for a class of 
quantum algorithms such as the Quantum Adiabatic Algorithm 
(QAA)~\cite{annealing}. However, the $H_P$ studied in~\cite{First} 
admit a single ground state (see~\cite{KS_new} for a recent discussion of 
this point), and therefore they do not show the complex 
clustering phenomenon explained above, whose influence on the spectrum of
the quantum problems remains unclear.

In this paper, we address this issue by studying the quantum version of a 
simplified model of random optimization problems introduced in~\cite{MZ08}, 
that shares part of the structure with more complex problems, 
at the same time being fully solvable in the thermodynamic limit. 

We find, for this model, that quantum fluctuations have a strong 
influence on the cluster structure:
they lift the degeneracy inside each cluster and the ground
state energy is lowered proportionally to the entropy of the cluster.
This mechanism is somehow similar
to the order-by-disorder phenomenon that is found 
in some frustrated magnets~\cite{OByD}.
Therefore, if the classical energy changes from cluster to cluster, 
then level crossings are induced between clusters by varying 
$\G$. These crossings happen over a continuous range of $\G$, giving rise to a complex 
spin glass phase characterized by a continuously changing ground state (as in 
classical spin glasses upon changing the temperature or the coupling 
constants~\cite{chaos})
and an exponentially small gap.
Finally, at large $\G$ the spin glass phase undergoes a first order ``delocalization'' transition towards 
a simple quantum paramagnetic phase, of the type discussed in~\cite{FirstMF,QREM,XORSAT}.
This set of results gives us a consistent picture of the effect of adding quantum fluctuations on top
of the complex classical phase space sketched in Fig.~\ref{fig1}, 
and allows to study in the same toy model the interplay between the
different phenomena discussed in~\cite{First} and \cite{FirstMF,QREM,XORSAT}.

\paragraph*{Model.---}
The quantum version of the Random Subcubes model~\cite{MZ08}
is defined as follows.
We take the Hilbert space $\HH$ of $N$ spins $1/2$ (qubits), 
in the basis of the Pauli matrices $\s^z_i$, 
$|\us \rangle = | \s_1, \cdots, \s_N \rangle$.
A {\it cluster} $A$ is a subset (subcube) of the Hilbert space 
$A = \{ |\us \rangle \, | \, \forall i, \s_i \in \pi_i^A \}$, where
$\pi^A_i$ are independent (set-valued) random variables which encode the
authorized values of $\s_i$:
$\pi^A_i = \{ -1 \}$ or $\{ 1 \}$ with probability $p/2$, and $\{-1,1\}$ 
with probability $1-p$.
The variable $i$ is ``frozen'' in $A$ in the former case and ``free'' in the latter.
Note that the number of states in a cluster $A$ is equal to $2^{Ns(A)}$, where $N s(A)$ 
is the number of free variables; we call $s(A)$ the {\it internal entropy} of 
a cluster.
We next define a set $\SS$ as the union of $2^{N (1-\a)}$ random clusters.
We define a Hamiltonian 
$H_A = N e_0(A) \sum_{\us \in A} | \us \rangle \langle \us |$
for each cluster, and
a ``penalty'' Hamiltonian
$H_V = N V \sum_{\us \notin \SS} | \us \rangle \langle \us |$.
The problem Hamiltonian $H_P = H_V + \sum_A H_A$ 
is then diagonal in the basis $|\us \rangle$.
We wish to interpret the states in $\SS$ as ``local minima''
of $H_P$ and the others as ``excited states''.
A sharp distinction between them
can be obtained by sending the positive constant $V$ to infinity;
for finite $V$, we will always assume that $V \gg \max_{A} e_0(A)$.
As a quantum term we choose here $\G H_Q = -\G \sum_{i=1}^N \s_i^x$, \ie a transverse field
acting on the spins. Note that taking instead an $H_Q$ proportional to 
$\sum_{\us,\us'}|\us \rangle \langle \us' |$
would lead to the simpler multi-solution Grover problem investigated 
in~\cite{Roland}.

\paragraph*{Analysis of the classical Hamiltonian.---}
In \cite{MZ08} it is shown that the set $\SS$ has the following structure.
{\it i)} For $\a \leq \a_{\rm d} = \log_2(2-p)$, each state $|\us \rangle$ belongs
to an exponential number of clusters and the space $\SS$ coincides with $\HH$.
{\it ii)} For $\a > \a_{\rm d}$, one has $\SS \neq \HH$. The number of clusters $\NN(s)$ of
entropy $s$ is given by
\beq
\Si(s) = N^{-1} \log_2 \NN(s) = 1 - \a - D(s || 1-p) \ ,
\eeq
where $D(x || y) = x \log_2(x/y) + (1-x) \log_2 [(1-x)/(1-y)]$. This
expression is restricted to the interval $s\in [s_{\rm min}, s_{\rm max}]$
for which $\Si(s) \ge 0$. 
Note that this dependency of the complexity $\Si(s)$ on the 
internal entropy is present in $k$-SAT and $q$-COL but not in XORSAT whose 
quantum version was studied in~\cite{XORSAT}.
Above $\a_{\rm d}$ there
is ``ergodicity breaking'' in the sense that a local random walk over solutions 
starting in one cluster takes an exponentially long (in $N$) time to reach 
another cluster~\cite{MZ08}.
{\it iii)} For $\a > \a_{\rm sep} = 1 + \log_2(1-p^2/2)/2$, the clusters are 
well separated, in the sense that with probability 1 for $N\to\io$ the Hamming
distance (number of opposed spins) between any two clusters is of order $N$.
From now on we focus on the region $\a > \a_{\rm sep}$, 
which is of principal interest for our purposes. 
Most results
are expected to hold more generically for $\a > \a_{\rm d}$ (because the 
effect of overlapping clusters is negligible~\cite{MZ08}), 
yet the arguments
used below should be refined in this regime

One can compute the total number of states in $\SS$ by observing that
\beq
| \SS | = 2^{N s_{\rm tot}} = \sum_A 2^{N s(A)} \sim \int_{s_{\rm min}}^{s_{\rm max}} {\rm d}s \, 2^{N [ \Si(s) + s]} \ ,
\eeq
therefore
$s_{\rm tot} = \max_{s\in[s_{\rm min},s_{\rm max}]} [ \Si(s) + s ]$.
It turns out that for $\a_{\rm d} < \a < \a_{\rm c} = p/(2-p) + \log_2 (2-p)$, the maximum is at 
$s^*(\a) \in [s_{\rm min}, s_{\rm max}]$.
Therefore most of the configurations of $\SS$ belong to one of the 
exponentially many (in $N$) clusters of size $s^*$. For $\a > \a_{\rm c}$,
the maximum is in $s^* = s_{\rm max}$, therefore most of $\SS$
is supported by the largest clusters whose number is sub-exponential
in $N$ since $\Si(s_{\rm max})=0$. The order parameter for this condensation
transition is the average Hamming distance between two random configurations
in $\SS$, close to $N/2$ for $\alpha<\alpha_{\rm c}$ and strictly 
smaller otherwise.
Finally, for $\a > 1$ there are no more clusters and the set $\SS$ is empty.

\paragraph*{Spectrum of the cluster Hamiltonian.---}
We will now study the spectrum of the quantum Hamiltonian $H = H_P + \G H_Q$ as
a function of $\G$, and 
we consider first the (``hard'')
$V\to \infty$ limit where $H_P$ is infinite for the states that do not 
belong to
$\SS$: then we can project out these states from the Hilbert space and look to the
restriction of $H = \sum_A H_A + \G H_Q$ on $\SS$, which contains $2^{N s_{\rm tot}}$ 
states.
Since the matrix $H_Q$ only connects configurations at unit Hamming distance, and
different clusters have distance of order $N$, the
Hamiltonian $H$ has no matrix elements connecting different clusters.
Therefore we can diagonalize $H$ separately in each cluster. 
The restriction of $H$ to a given cluster $A$
with $N s(A)$ free spins is equal to $H_A$ plus the Hamiltonian of $N s(A)$ 
uncoupled spins in a transverse
field, its spectrum is hence made of levels 
\beq
E_k(A) = N e_0(A) + (2 k - N s(A) ) \G \ , 
\label{eq_Ek}
\eeq
with $k = 0,\cdots,Ns(A)$,
each $\binom{N s(A)}{k}$
times degenerate. In particular the lowest level has energy per spin
$e_{\rm GS}(A) = e_0(A) - \G s(A)$, therefore
the energy of clusters with larger entropy 
decreases faster with $\G$.

\paragraph*{Quantum paramagnetic state.---} 
Next, we consider a ``soft'' version of the model in which $V$ is finite
(still with $V \gg \max_{A} e_0(A)$).
Therefore now $H$ is defined on the full Hilbert space $\HH$.
In this case, in addition to the $2^{N s_{\rm tot}}$ energy levels discussed above
(that we shall refer to as the $\SS$-band), 
there exists another set
of $2^{N} - 2^{N s_{\rm tot}} \sim 2^{N}$ levels (the $V$-band), whose energy is
expected to be of order $V$ at small $\G$.
For the states in the $\SS$-band we use perturbation theory in $\G$; the leading
order is of the form of Eq.~(\ref{eq_Ek}), and at any finite order $n$ the correction is
$O((\G^{2}/(NV))^n)$, so it vanishes in the thermodynamic limit.
Moreover, an argument based on small rank perturbation analysis~\cite{smallrank}
shows that the spectrum of the $V$-band states is close to the one of $N$ free spins 
in transverse field with classical energy~$N V$:
\beq
E_k^V = N V + (2 k - N ) \G \ , \ \ \ k = 0,\cdots,N \ ,
\eeq
with degeneracy close but not equal to $\binom{N}{k}$.
In particular the lowest of such levels is the 
{\it Quantum Paramagnetic (QP) state} $| QP \rangle \sim 2^{-N/2} \sum_{\us} |\us \rangle$,
which is uniformly extended in the basis 
$| \us \rangle$ and has energy per spin
$e_{\rm QP} = V - \G$.

\begin{figure}
\includegraphics[width=.47\textwidth]{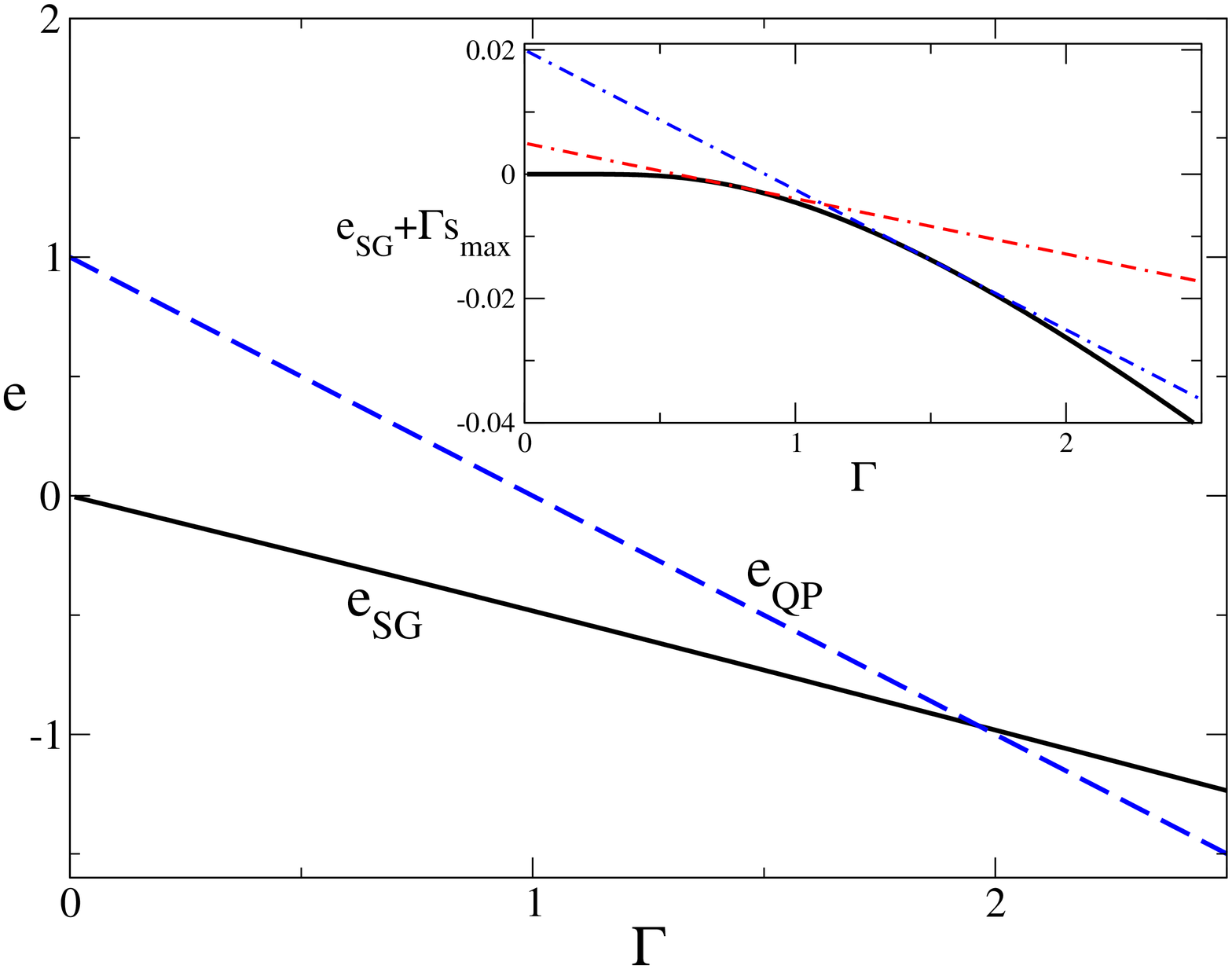}
\vskip-10pt
\caption{
Low energy spectrum of the model for $\a_{\rm sep}<\a<\a_{\rm s}$.
As an example we choose (following~\cite{MZ08}) $p=0.7$, $\a=0.85$, and
$\Si(e_0) = (1-\a)[2 + e_0/e_{\rm m} - (e_0/e_{\rm m}) \ln(e_0/e_{\rm m})]/3$
for $e_0 \in [0,e_{\rm m}]$ with $e_{\rm m}=0.1$.
{\it Main panel}: Energy of the SG (cluster) ground state
[Eq.~(\ref{SGe}), full line]
and of the QP state $e_{\rm QP} = V - \G$ 
for $V=1$ (dashed line).
A first order transition between the two states happens at $\G \sim 2$.
{\it Inset}: Level crossings in the SG state.
For better readability we plot $e_{\rm SG}+\G s_{\rm max}(0)$ [Eq.~(\ref{SGe}), full line] 
and show the energy
$e_0 - \G [s_{\rm max}(e_0) - s_{\rm max}(0) ]$ of two different clusters
with $e_0 = 0.05, 0.2$ (dot-dashed lines).
}
\vskip-15pt
\label{fig3}
\end{figure}

\paragraph*{Level crossings.---}
We discuss now the zero temperature phase diagram of the model
for $\a > \a_{\rm sep}$ and $N \to \io$. To get a meaningful
thermodynamic limit,
the number of clusters of energy $e_0$ is
set to $2^{N \Si(e_0)}$, where $\Si(e_0)$ is some increasing function 
of $e_0 \in [0,e_{\rm m}]$
(as in most random optimization problems).
We assume that $\Si(e_{\rm m}) = 1-\a$ so the total number of 
clusters in $\SS$ is still $2^{N(1-\a)}$.
Since the frozen variables are chosen independently
for each cluster, the complexity of clusters of energy 
$e_0$ and entropy $s$ is
$\Si(e_0,s) = \Si(e_0) - D(s || 1-p)$.
It vanishes for a given value $s_{\rm max}(e_0)$ which is also an increasing function of $e_0$.
The $\SS$-band, or {\it Spin Glass (SG)},
ground state energy is
\beq\begin{split}\label{SGe}
e_{\rm SG} &= \min_{e_0 \in [0,e_{\rm m}]} 
\left[ \min_{s\in [s_{\rm min}(e_0),s_{\rm max}(e_0)]} 
(e_0 - \G s) \right]
\\ & =\min_{e_0\in [0,e_{\rm m}]}\big[ e_0 - \G s_{\rm max}(e_0) \big] \ .
\end{split}\eeq
The minimum is in $e_0 =0$ as long as $\G < \G_{\rm lc} =1/(s_{\rm max}'(0))$. Above this value,
the minimum is in a different $e_0$ for each value of $\G$: in this region the ground state
changes abruptly from one cluster to another upon changing $\G$ by an infinitesimal amount~\cite{chaos}, see the inset of Fig.~\ref{fig3}.
Since the clusters have Hamming distance proportional to $N$, we expect these crossings
to be avoided at finite $N$ producing exponentially (in $N$) small gaps~\cite{QREM,First}.
Note that in some relevant cases the slope of $\Si(e_0)$ in $e_0=0$ is infinite, therefore
$\G_{\rm lc} =0$ and level crossings happen at all $\G$.

The energy $e_{\rm QP}$ crosses
$e_{\rm SG}$ given by Eq.~(\ref{SGe}), giving rise to a first order phase
transition between the spin glass and the quantum paramagnet~\cite{QREM,FirstMF,XORSAT} at
a critical $\G \propto V$, see Fig.~\ref{fig3}. As a consequence, the transverse magnetization $m_x = de/d\G$
has a jump at the transition~\cite{XORSAT}.

\begin{figure}
\includegraphics[width=.47\textwidth]{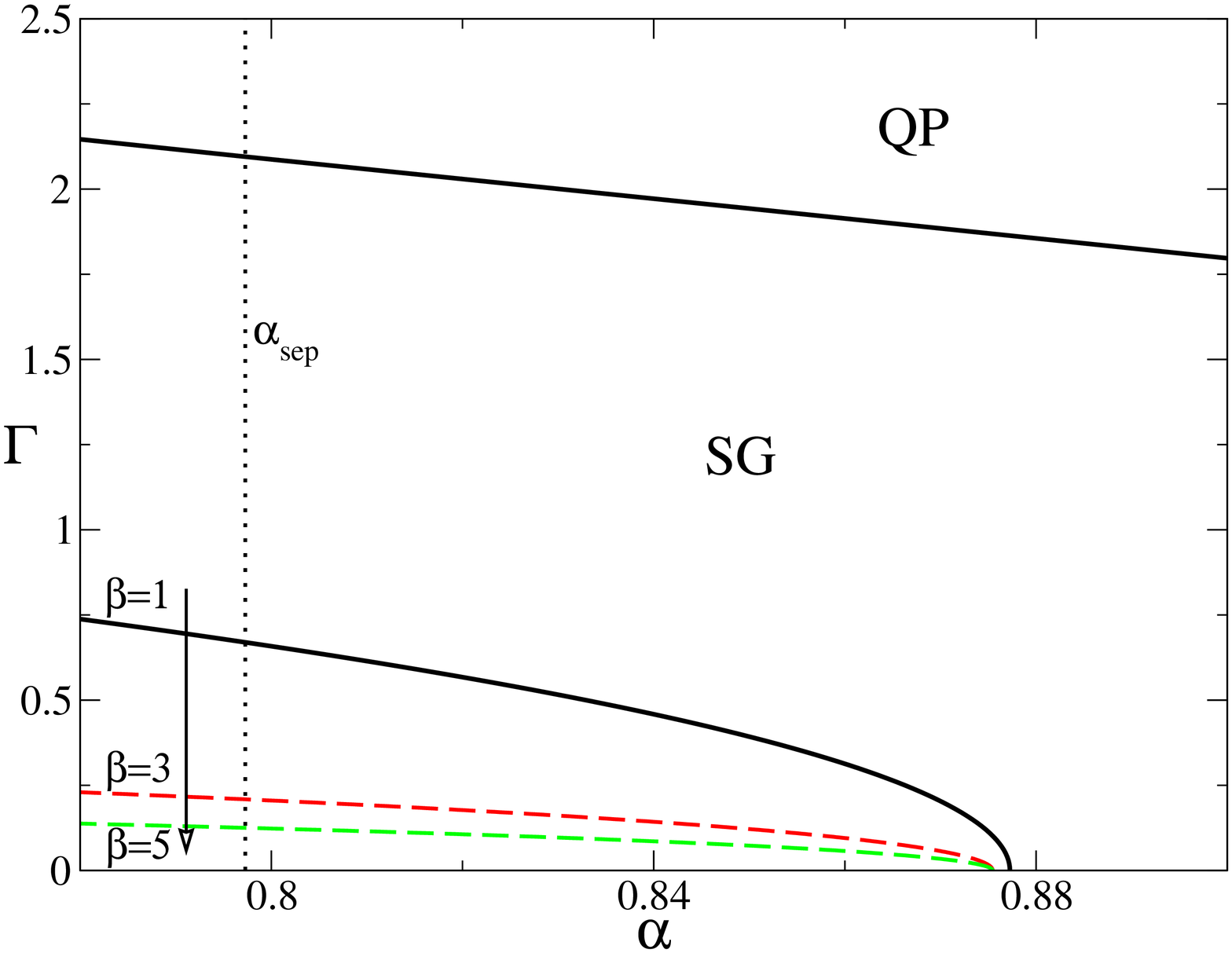}
\vskip-14pt
\caption{
Phase diagram of the model for $p=0.7$,
$\Si(e_0)$ as in Fig.~\ref{fig3}, and $\b=1$ (full lines).
The vertical line corresponds to $\a_{\rm sep}=0.797$ for this value of $p$.
The higher $\G$ line is the first order transition between SG and QP.
Above the lower $\G$ line $\a_{\rm c}(\G,\b=1)$ the system is in the condensed
phase.
The condensation transition lines $\a_{\rm c}(\G,\b)$ are also reported (dashed lines) for 
different values of $\b$, showing that the non-condensed phase disappears
for $\b \to \io$.
The complexity of the zero-energy clusters is $\Sigma(e_0=0)=2(1-\a)/3$, hence
one has $\a_{\rm c}(\G=0,\b=\io)=\frac{2p-1}{2-p} + \frac32 \log_2(2-p) = 0.875$.
}
\vskip-15pt
\label{fig2}
\end{figure}

\paragraph*{The condensation transition.---}
The previous analysis shows that in the region $\a_{\rm sep} < \a < \a_{\rm c}$ the perturbation
$\G H_Q$ has a dramatic effect. At $\G =0$, most of the states in $\SS$ belong to one of exponentially
many (in $N$) small clusters, while at any $\G > 0$ the few largest clusters of entropy $s_{\rm max}$ have
the smallest energy. A more complete picture is obtained by studying the model at finite
temperature. It is convenient to separate the contribution of the two phases to the partition
function, $Z = \Tr \, e^{-\b H} = Z_{\rm SG} + Z_{\rm QP}$,
with $c = 2 \cosh(\b \G)$:
\begin{eqnarray}
Z_{\rm QP} &\sim& \sum_k e^{-\b E_k^V} = e^{-\b N V} c^{N} \ , \\
Z_{\rm SG} &\sim&
\sum_{A,k} e^{-\b E_k(A)} = 
\int {\rm d}e_0 {\rm d}s \, 2^{N \Si(e_0,s)} e^{-\b N e_0}
c^{N s} \ .\nonumber
\end{eqnarray}
The free energy is $f = -(T/N) \ln Z = \min\{f_{\rm SG}, f_{\rm QP}\}$, 
with $f_{\rm QP} = V - T \ln c$ and
\beq\label{fSG}
f_{\rm SG} = - T \hskip-15pt \underset{
\substack{e \in [0,e_{\rm m}] \\ s\in[s_{\rm min}(e),s_{\rm max}(e)]}
} 
{\max}
\hskip-10pt
[ \Si(e_0,s) \ln 2 - \b e_0 + s \ln c] \ .
\eeq
The first order transition happens when the free energies cross, while
the condensation transition $\a_c(\G,\b)$ happens when the maximum in Eq.~(\ref{fSG}) is attained in $s_{\rm max}$ 
for the first time; these lines are plotted in Fig.~\ref{fig2}. 
We observe that in the limit $\b \to \io$, the lines $\a_c(\G,\b)$ shrink to the 
horizontal axis and the system is in the condensed phase 
for any~$\G~>~0$. The first order transition to the QP phase happens
for larger values of $\G$ and is almost independent of $\b$ at low temperature (for $\b\gtrsim 1$).

\paragraph*{Conclusions.---}
In this paper we introduced the quantum version of a simple toy model of optimization 
problems~\cite{MZ08}.
In the classical case $\G=0$, the model captures the essential
structure of the space of solution of random optimization problems, and
displays several phase transitions
that are present also in more realistic problems such as $k$-SAT and $q$-COL,
at least at large $q,k$. 
We explored the consequences of this complex
structure on the spectrum of the quantum Hamiltonian at $\G>0$, and we showed that:
{\it i)} Quantum fluctuations lower the energy of a cluster proportionally to
its size.
{\it ii)} As clusters have an energy distribution,
level crossing between different clusters
are induced as a function of $\G$ in the spin glass phase, due
to a competition between energetic and entropic effects.
These crossings
happen in a continuous range of $\G$, giving rise to a complex spin glass 
phase characterized by a continuously changing
ground state. Since the clusters are separated by an extensive Hamming distance,
as were individual solutions considered in~\cite{First}, we expect 
an exponentially (in system size) small gap
everywhere in this phase.
{\it iii)} At large $\G \sim V$ the spin glass phase undergoes 
a first order transition towards a
quantum paramagnetic phase~\cite{FirstMF,QREM,XORSAT},
corresponding to the complete delocalization of the
ground state in the computational basis $| \us \rangle$.
{\it iv)} At finite temperature, there is a line of condensation 
transitions $\a_{\rm c}(\Gamma,\b)$ that shrinks to $\Gamma=0$ at low temperatures:
indeed, at zero temperature 
the condensation transition becomes abrupt. 
While at $\G=0$ the space of solutions is dominated by an exponential number of 
clusters of intermediate size, for any $\G > 0$ the biggest
clusters contain the ground states.

Overall, this toy model shows that the low energy spectrum
of quantum optimization problems can be very complex, and characterized by different level crossings:
internal level crossings in the spin glass phase, or the crossing between the spin glass and the quantum
paramagnet giving rise to a first order phase transition. Moreover, both entropic and energetic effects
are important. We expect that this complex structure of the low-energy spectrum of the quantum Hamiltonian will
have deep consequences on the behavior of quantum algorithms: for instance, the Quantum Adiabatic
Algorithm proposed in~\cite{annealing} should run into difficulties because of the exponentially
small gaps that are expected at the crossings.

We expect these results to be also relevant for the physics of quantum glasses: indeed, mean-field lattice glass models
fall in the same class of random optimization problems. In this context the glass transition corresponds to
the condensation transition discussed above.
The phase diagram we reported in Fig.~\ref{fig2}, characterized by a
re-entrance of the glass transition line, 
has been found using different approaches~\cite{DR}.
Extending these results to more realistic optimization problems, and
further investigating the connection between the spectrum and the performances of quantum algorithms,
are two lines of research that should be developed in the future.

\acknowledgments{
We wish to thank B.~Altshuler, G.~Biroli, F.~Krzakala, R.~Monasson,
D.~Reichman, J.~Roland, G.~Santoro and L.~Zdeborov\'a for important discussions. 
}

\vskip-17pt


\end{document}